\documentclass[10pt]{article}
\usepackage{amsfonts}
\usepackage{amssymb}
\usepackage{amsthm}

\usepackage{graphicx}

\newcommand{\sss}{\setcounter{equation}{0}}
\newtheorem{theorem}{THEOREM}[section]

\newtheorem{remark}[theorem]{REMARK}

\newtheorem{prop}[theorem]{PROPOSITION}

\newtheorem{assumption}[theorem]{ASSUMPTION}

\newcommand{\ere}{ {\mathbb R}}

\newcommand{\ese}{{\mathbb S}}

\def\beq{\begin{equation}}
\def\ene{\end{equation}}
\def \ds {\displaystyle}
\newcommand{\bull}{\hfill $\Box$}
\def\x{\mathbf x}
\def\p{\mathbf p}
\def\q{\mathbf q}
\def\e{\hbox{\rm erf}}

\begin{document}
\baselineskip=20 pt
\parskip 6 pt

\title{Entanglement Creation in Low-Energy Scattering
\thanks{ PACS Classification (2010): 03.67.Bg, 03.65.Nk, 03.65.Db.  Mathematics Subject Classification(2010):81P40, 81U05, 35P25.}
\thanks{ Research partially supported by
 CONACYT under Project CB-2008-01-99100.}}
 
 \author{ Ricardo Weder  \thanks { On Leave of absence from Instituto de Investigaciones en Matem\'aticas Aplicadas y en
 Sistemas. Universidad Nacional Aut\'onoma de M\'exico. Apartado Postal 20-726,
M\'exico DF 01000.}\thanks {Fellow, Sistema Nacional de Investigadores.
Electronic mail: weder@unam.mx  }
\\
 Institut National de Recherche en Informatique et en Automatique
Paris-Rocquencourt.\\
 Projet POEMS. Domaine de Voluceau-Rocquencourt, BP 105,
78153, Le Chesnay Cedex  France.}

\date{}
\maketitle


\vspace{.5cm}
 \centerline{{\bf Abstract}}
\noindent
 We study the entanglement creation in the low-energy scattering of two particles  in three dimensions, for  a general class of interaction  potentials that are not required to be spherically symmetric. The incoming asymptotic state, before the collision, is a product of two normalized Gaussian states.  After the scattering the particles are  entangled.  We take as a measure of the entanglement the purity of one of them. We provide a rigorous explicit computation, with error bound, of the leading  order of the  purity at  low-energy.  The entanglement  depends strongly in the difference of the masses. It takes its minimum when the masses are equal, and  it increases rapidly with the difference of the masses. 
  It is quite remarkable that the anisotropy of the potential gives no contribution to the leading order of the purity,  on spite of the fact that entanglement  is a second order effect.

\bigskip
\noindent 
\section{Introduction}\sss
In this paper we study how entanglement is created in a scattering process. This topic  has intrinsic
interest. As is well known,  scattering is a basic dynamical process that is essential across all areas of physics. Entanglement is a central notion of modern quantum theory, in particular, it is the fundamental resource for quantum information theory and quantum computation. It is a measure for quantum correlations between subsystems. In the case of bipartite systems in pure states, entanglement is a measure of  how far away from being a product state a  pure state of the  bipartite system is.  Product states are  called disentangled.  It is now well understood that entanglement in a pure bipartite quantum state  is equivalent to the degree of mixedness of each subsystem. See, for example  \cite{nc}, \cite{hh}, \cite{ai}. Moreover, the study of  entanglement creation  in scattering is interesting  for a many other
reasons. For example, for the implementation of quantum information processes in physical systems where scattering is central to the dynamics, like  ultracold atoms and  solid state devices. Moreover, the study of entanglement   in the scattering of particles  requires   quantum information theory with continuous variables and mixed continuous-discrete variables. See \cite{ai} for a review of this topic. As scattering interactions are fundamental at all scales, and as there is a large variety of scattering systems, it is possible that scattering will provide a new perspective to  quantum information theory.  Finally, entanglement creation is important to the theory of scattering itself, because it poses new problems that can shed some new light and new points of view in the study of scattering processes.

Actually,  from the conceptual point of view  scattering is perhaps  the simplest   way to entangle two particles.
Before the scattering, in the incoming state,  the two particles are in a pure product state where they are uncorrelated. As they approach each other they become entangled  by sharing quantum information between them. After the scattering, when they are far apart from each other, they  remain entangled  in the outgoing asymptotic state, that is not a product state anymore.

We  take as measure of entanglement of a pure state the purity of one of the particles, that is to say, the trace of the square of the reduced density matrix of one of the particles, that is obtained by taking the trace on the other particle of  the density matrix of the pure state. The purity of a product state is one. 
  
  We consider two spinless particles in three dimensions with  the interaction given by a general potential that is not required to be spherically symmetric. Initially the particles are in an incoming asymptotic state that is a product of two Gaussian states. After the scattering the particles are in an outgoing asymptotic state that, as mentioned above, is not a product state and our problem is to determine the loss of purity of one of the particles, due to the entanglement with the other that is produced by the scattering process.

The Hilbert space of states for the two particles in the configuration representation is  $\mathcal H:= L^2(\ere^6)$. The Schr\"odinger equation is
\beq\label{1.1}
i \hbar \frac{\partial }{\partial t}\varphi(\x_1,\x_2)= H\varphi(\x_1.\x_2),
\ene 
where the Hamiltonian is given by
\beq \label{1.2}
H=H_0+V(\x_1-\x_2),
\ene
with $H_0$ is the free Hamiltonian,
\beq\label{1.2b}
H_0:= -\frac{\hbar^2}{2m_1} \Delta_1 -\frac{\hbar^2}{2m_2} \Delta_2,
\ene
where  $\hbar$  is Planck's constant, $m_j,  j=1,2$, are, respectively, the mass of particle one and two, and  $\Delta_j$, the Laplacian in the coordinates $\x_j, j=1,2$. The potential of interaction is multiplication by a real-valued function, 
$V(\x)$, defined for $ \x\in\ere^3$. As usual,  we assume that the interaction depends on the difference of the coordinates   
  $\x_1-\x_2$, but no spherical symmetry is  supposed. We assume that $V$ satisfies mild assumptions on its regularity and its decay at infinity. See Assumption \ref{ass2.1} in Section 2. For example, $V(\x)$ satisfies Assumption  \ref{ass2.1} if there are constants $R, C>0$ such that,

\beq \label{1.3}
\int_{|\x| \leq R} |V(\x)|^2 d\x <\infty,
\ene
and
\beq \label{1.4}
|V(\x)| \leq C (1+|\x|)^{-\beta'}, \quad \hbox{\rm for}\, |\x| \geq R,
\ene
for some $\beta' > \beta$, with $\beta$ as in Assumption  \ref{ass2.1}. Note that $\beta$ controls the decay rate of the potential at infinity. Remark  that (\ref{1.3}) allows for  Coulomb local singularities. We also suppose that at zero energy there is neither an eigenvalue nor a resonance (half-bound state), what generically is true. See Section 2. 

We work in the  center-of-mass frame and we consider an incoming asymptotic state that is a product of two normalized Gaussian states, given in the momentum representation by,
\beq\label{1.5}
\varphi_{\rm in,\p_0}(\p_1,\p_2):=  \varphi_{\p_0}(\p_1) \,  \varphi_{-\p_0}(\p_2),
\ene
with,
\beq\label{1.6}
\varphi_{\p_0}(\p_1):=  \frac{1}{(\sigma^2\pi)^{3/4}}   e^{- (\p_1-\p_0)^2 /  2\sigma^2},
\ene
where $\p_i,i=1,2$ are, respectively, the momentum of particles one and two. 

In the state (\ref{1.5}) particle one has mean momentum $ \p_0$ and particle two has  mean momentum $ -\p_0$.
 The variance of the momentum distribution of both particles is $\sigma$. We assume that the scattering takes place at the origin at time zero, and for this reason  the   average position of both particles  is zero in the incoming asymptotic state   (\ref{1.5}). 
After the scattering process is over the two particles are  in the outgoing  asymptotic state, $\varphi_{\rm out, \p_0}$, given by
 
\beq\label{1.7}
\varphi_{\rm out,\p_0}(\p_1,\p_2): = \left(  \mathcal S(\p^2/2m)  \varphi_{\mathrm in, \p_0 }\right)(\p_1,\p_2),
\ene
where $\p:= \frac{m_2}{m_1+m_2}\p_1-\frac{m_1}{m_1+m_2}\p_2$ is the relative momentum, $m:= m_1\,  m_2/(m_1+m_2)$ is  the  reduced mass, and $ \mathcal S( \p^2/2m )$ is the scattering matrix for the relative motion.

The purity of  $\varphi_{\rm out, \p_0}$ is given by,
\beq\label{1.8}
{\mathcal P}(\varphi_{\rm out, \p_0})= \int_{\ere^{12}}\, d\p_1 \,d\p_1' \,d\p_2 \,d\p_2' \,\varphi_{\rm out, \p_0}(\p_1,\p_2)
 \,\overline{\varphi_{\rm out, \p_0}(\p_1', \p_2)} \,\varphi_{\rm out, \p_0}(\p_1',\p_2') \,\overline{\varphi_{\rm out, \p_0}(\p_1,\p_2')}.
\ene
Since the relative momentum, $\p$, depends on $\p_1$ and on $\p_2$, 
$\varphi_{\rm out, \p_0}$ is no longer a product state and it has purity smaller than one, what means that entanglement between the two particles has been created by the scattering process. 

Observe that in the state (\ref{1.5}) the mean relative momentum of the particles  is  equal to $\p_0$.

Note that to be in the low-energy regime  we need that the mean relative momentum $\p_0$ be small, but also that the variance $\sigma$ be small, because if $\sigma$ is large, the incoming asymptotic state   $\varphi_{\rm in, \p_0}$ will have a big probability of having large momentum, even if the mean relative momentum $\p_0$ is small.  

We denote by $\varphi_{\rm in}$  the incoming asymptotic state with mean relative momentum $\p_0=0$, and we designate, $\varphi_{\rm out}:=   \mathcal S(\p^2/2m)  \varphi_{\mathrm in} $. 

We denote by
$$
 \mu_i:= \frac{m_i}{m_1+m_2}, i=1,2,
$$
the fraction of the mass of the $i$ particle to the total mass.

In Theorems \ref{theor3.2} and \ref{theor3.6}  in Section 3 we give a rigorous proof of the following results  on the leading order of the purity at low energy.  

\beq\label{1.9}
 \mathcal P(\varphi_{\rm out, \p_0})= \mathcal P(\varphi_{\rm out})+O(|\p_0/\hbar|), \hbox{\rm as}\, |\p_0/\hbar| \rightarrow 0,
 \ene
 where  $O(|\p_0/\hbar|)$ is uniform on $\sigma$, for $\sigma$ in bounded sets.
Furthermore, with $\beta$ as in Asumption  \ref{ass2.1} (recall   that $\beta$ controls the decay rate of the potential at infinity),

\beq\label{1.10}
 \mathcal P(\varphi_{\rm out})=1- ( c_0\sigma/ \hbar)^2 \mathcal E(\mu_1) +\left\{
 \begin{array}{clcr}
o\left( |\sigma/\hbar|^2\right),    & \hbox{\rm if}\, \beta >5, \\
  O\left(|\sigma/\hbar|^3 \right),   &   \hbox{\rm if}\, \beta >7,
  \end{array}
 \right.
 \ene
  with $c_0$ the scattering length that is defined in (\ref{2.20b}) and   where the entanglement coefficient $\mathcal E(\mu_1)$ is given by,
\beq\label{1.11}
\mathcal E(\mu_1):=  \frac{16}{\pi\, \left( 1+ (2\mu_1-1)^2 \right)}+ \frac{4}{ (2\mu_1-1)^2}\, \frac{\left(1+(2\mu_1-1)^2\right)^{3/2}-1 }{\sqrt{1+(2\mu_1-1)^2}}-8
J(\mu_1,1-\mu_1)-8 J(1-\mu_1,\mu_1),
\ene
with
\beq \label{1.12}\begin{array}{c}
J(\mu_1,\mu_2):= \frac{1}{\pi^{9/2}}\, \int d\q_2\left[ \,\,\int d\q_1 |\mu_2\q_1-\mu_1\q_2|\,\, \hbox{\rm Exp}[-\frac{1}{2}(\mu_1^2+\mu_2^2)(\q_1+\q_2)^2-(\mu_2\q_1-\mu_1\q_2)^2 - \q_1^2/2] \right.\\\\\left.
\ds \frac{\sinh[(\mu_1-\mu_2)\, |\q_1+\q_2|\,|\mu_2\q_1-\mu_1\q_2|]}{(\mu_1-\mu_2)\, |\q_1+\q_2|\,|\mu_2\q_1-\mu_1\q_2|}\,\,\right]^2.
 \end{array}
\ene
 In the appendix we explicitly evaluate $J(1/2,1/2)$ and $J(1,0$),
 \beq\label{1.13}
 J(1/2,1/2)= \frac{3}{2}+\frac{1}{\pi} \, \left(\frac{\sqrt{27}}{4}-3 \arctan\left(\frac{1}{2-\sqrt{3}}\right)\right) =0.663497,
\ene

\beq \label{1.14}
J(1,0)= 2(1+ \frac{1}{\sqrt{3}} -\sqrt{2})= 0.32627.
\ene 
For  $\mu_1 \in [0,1]\setminus\{1/2,1\}$ we compute  $J(\mu_1,1-\mu_1)$ numerically using Gaussian quadratures. 

Observe that $\mathcal E(\mu_1)= \mathcal E(1-\mu_1)$,  as it should be, because $\mathcal P(\varphi_{\rm out})$ is invariant under the exchange of particles one and two.

Note that there is no term of order $\sigma/\hbar$ in (\ref{1.10}). Actually,  the terms
 of order  $\sigma/\hbar$ cancel each other because of the unitarity of the scattering matrix. This shows that for low energy the entanglement is a second order effect. 
 
 The scattering length $c_0$ is a measure of the strength of the interaction. As is well known, and can be seen in 
Theorem \ref{th2.2} in Section 2, at first order  for low energy the scattering is isotropic and the total cross section, that is given by $4 \pi c_0^2$,  is determined by the scattering length, $c_0$. However, the effects of the anisotropy of the potential appear at second order. It is quite remarkable that these effects give no contribution to  the evaluation of the leading order of the purity. It follows from this that  the leading order of the entanglement for low energy  (\ref{1.10}) is determined by the scattering length, $c_0$,  and that  the anisotropy of the potential plays no role, on spite of the fact that entanglement  is a second order effect, what is surprising.

We see from Table 1  and Figure 1 that the entanglement  coefficient depends strongly in the difference of the masses. It takes its minimun for $\mu_1=0.5$, when the masses are equal, and  it increases rapidly with the difference of the masses, as $\mu_1$ tends to one.  This shows that, if the scattering length is fixed,   the entanglement  takes its minimum when the masses are equal and that it strongly increases with the differences of the masses.  This is indeed a remarkable result. Suppose that we consider different pairs of particles that interact in the same way at low energy, in the sense that, to leading order, they have the same total scattering cross section, i.e., such that the scattering length, $c_0$, is the same for all the  pairs. Moreover, suppose that the total mass, $M$, of the pairs is keep fixed, but that the individual masses, $m_1,m_2$ of the particles are different in each pair. Our results show that, under  these conditions, over four times more entanglement is produced by increasing the difference of the masses of the particles in the pairs. In practical terms, this means that in experimental devices to produce entanglement by scattering processes it is advantageous to use particles with a large mass difference.

This fact can be understood in a  physically intuitive   way as follows: in the scattering of  a light particle with  a very heavy one, the trajectory of the light particle will be strongly changed, with a large exchange of quantum information between the particles, leading to a large entanglement creation.

 Note that in the scattering of a particle with a large mass and a particle with a small mass we can assume that the trajectory of the large particle is not affected by the interaction, i.e. that, to a good approximation, it follows a free trajectory, and that the small particle feels a  (external) interaction potential centered in the position of the large particle. However,  the trajectory of the small particle will be strongly affected by the interaction, what will produce exchange of information between the particles, leading to the creation of entanglement between them.  To evaluate this entanglement it is, however, necessary to take into account the degrees of freedom of both particles, as we do to compute the purity.

In the paper \cite{wlc} a similar problem is considered in the case of equal masses and spherically symmetric potentials. They  give an approximate expression for the leading order of the purity in the case of  a Gaussian incoming wave packet that is very narrow in momentum space.

The generation of entanglement in scattering processes  has been previously considered in one dimension, mainly for potentials with explicit solution. See \cite{sj}, \cite{hs}, and the references quoted there. Moreover,  \cite{da},  \cite{dfa},  \cite{afft}, and  the references quoted there, consider a system of heavy and light particles. They study the asymptotic dynamics and  the decoherence produced on the heavy particles  by the scattering with light particles in the limit of   small mass ratio, what is  different from our problem. The loss of quantum coherence induced on  heavy particles by the interaction with light ones  has attracted much interest. See for example \cite{jz}, and \cite{gf}.

The paper is organized as follows. In Section 2 we  define the wave and scattering operators, the scattering matrix and we consider its low-energy behavior. In Section 3 we prove our results in the creation of entanglement. In Section 4 we give our conclusions. In the Appendix we compute integrals that we need in Section 3. Along the paper we denote by $C$ a generic positive constant  that does not necessarily have the same value in different appearances. 
\section{Low-Energy Scattering}
\sss

We consider the scattering of two spinless particles in three dimensions. We find it convenient to use the time-dependent formalism of scattering theory. See, for example, \cite{ya1,ya2,we,tay, ne}.

The Hilbert space of states in the configuration representation is  $\mathcal H:= L^2(\ere^6)$. The Schr\"odinger equation is
\beq\label{2.1}
i \hbar \frac{\partial }{\partial t}\varphi(\x_1,\x_2)= H\varphi(\x_1.\x_2),
\ene 
where the Hamiltonian is given by
\beq \label{2.2}
H=H_0+V(\x_1-\x_2).
\ene
The operator $H_0$ is the free Hamiltonian,
\beq\label{2.3}
H_0:= -\frac{\hbar^2}{2m_1} \Delta_1 -\frac{\hbar^2}{2m_2} \Delta_2,
\ene
with $\hbar$ Planck's constant, $m_j, j=1,2$, respectively, the mass of particle one and two, and   $\Delta_j$ the Laplacian in the coordinates $\x_j, j=1,2$. The potential of interaction is multiplication by a real-valued function, $V(\x)$, defined for $ \x\in\ere^3$. As usual,  we assume that the interaction depends on the difference of the coordinates  $\x_1-\x_2$, but no spherical symmetry is  supposed. $V$ satisfies the following condition.

\begin{assumption}\label{ass2.1}
{\rm For some $\beta >0$,   $(1+|x|)^\beta V(\x)$     is a compact operator from the Sobolev space
 $H^1$ into the Sobolev space  $H^{-1}$ . 
}
\end{assumption} 

 Below we will assume that $ \beta > 5$ or that $\beta >7$. For the definition of Sobolev's spaces see \cite{ad}. Conditions for  Assumption 2.1 to hold are well know \cite{sch}, \cite{kj}. For example, if (\ref{1.3}, \ref{1.4}) are satisfied. 
 
 Under this condition $H$ is defined  as  the quadratic form sum of $H_0$ and $V$ and it is a  self-adjoint operator.
 
 The wave operators are defined as
 $$
 W_\pm:= \hbox{\rm s-lim}_{\pm \infty}\, e^{i \frac{t}{\hbar}H} \, e^{ -i \frac{t}{\hbar}H_0}.
 $$
 As is well known, under our condition the wave operators exist and are asymptotically complete, i.e. their ranges coincide with the absolutely continuous subspace of $H$. Moreover, the scattering operator,
 \beq \label{2.4}
 S= W_+^\ast \, W_-,
 \ene 
 is  unitary.
 
Before the scattering, when  the  particles are far apart  from each other and the interaction is weak, the dynamics of the system is well approximated by an incoming  solution to the free Schr\"odinger equation with the potential set to zero,

$$
e^{ -i \frac{t}{\hbar}H_0} \, \varphi_-,
$$
where the incoming asymptotic state $\varphi_-$ is the Cauchy data at time zero  of the incoming  solution to the free Schr\"odinger equation.  When the particles are close to each other, and the potential is strong, the dynamics of the system is 
given by the solution to the Schr\"odinger equation,
\beq\label{2.4b}
e^{ -i \frac{t}{\hbar}H} \, W_-\varphi_-,
\ene
 that is asymptotic to the incoming solution to the  free Schr\"odinger equation as $ t\rightarrow -\infty$,
 $$
 \lim_{t \rightarrow -\infty} \left\|   e^{ -i \frac{t}{\hbar}H_0} \, \varphi_- - e^{ -i \frac{t}{\hbar}H} \, W_-\varphi_-\right\|=0.
 $$
 After the scattering, for large positive times, the particles again are far away from each other and the dynamics of the system is 
  well approximated by the  outgoing  solution to the free Schr\"odinger equation
  $$
  e^{ -i \frac{t}{\hbar}H_0} \, W_+^\ast\,W_-\varphi_-,
  $$
 that is asymptotic to the solution to the Schr\"odinger equation (\ref{2.4b}) as $t \rightarrow \infty$,
 
 $$
 \lim_{t \rightarrow \infty} \left\| e^{ -i \frac{t}{\hbar}H} \, W_-\varphi_- - e^{ -i \frac{t}{\hbar}H_0} \,  W_+^\ast\,W_-\varphi_-
 \right\|=0.
 $$
 The outgoing asymptotic state is the Cauchy data at time zero of the  outgoing  solution to the free Schr\"odinger equation:  $\varphi_+:=W_+^\ast\,W_-\varphi_-$. It is given by the scattering operator, $\varphi_+= S \varphi_-$.

As usual, we consider the center-of-mass and relative distance coordinates,
\begin{eqnarray}\label{2.5}
\x_{\rm cm}:= \frac{m_1\x_1+m_2\x_2}{m_1+m_2}, \\
\x:= \x_1-\x_2.
\end{eqnarray}

The state space $\mathcal H$ factorizes under this change of coordinates as,
\beq \label{2.6}
\mathcal H= {\mathcal H}_{\rm cm} \otimes {\mathcal H}_{\rm rel},
\ene
Where ${\mathcal H}_{\rm cm}=L^2(\ere^3), {\mathcal H}_{\rm rel}:=L^2(\ere^3)$ are, respectively, the state spaces for the center-of-mass motion and the relative motion. Since the interaction depends only on $\x$, the Hamiltonian and  the wave and  scattering operators decompose under the tensor product structure and, in particular, we have that
\beq\label{2.7}
S= I_{\rm cm}\otimes S_{rel},
\ene
where $I_{\rm cm}$ is the identity  operator on ${\mathcal H }_{\rm cm}$ and $S_{\rm rel}$ is the scattering operator for the relative motion, that is defined as follows. The Hamiltonian for the relative motion is given by,
\beq\label{2.8}
H_{\rm rel}:= - \frac{\hbar^2}{2m} \Delta_{\x} +V(\x),
\ene
where $m$ is the reduced mass,
\beq\label{2.9}
m:= \frac{m_1 \, m_2}{m_1+m_2},
\ene
and $\Delta_{\x}$ is the Laplacian in the $\x$ coordinate. The free relative Hamiltonian is,
\beq\label{2.10}
H_{0, \rm rel}:= - \frac{\hbar^2}{2m} \Delta_{\x}. 
\ene

The relative wave operators are defined as,
 \beq\label{2.11}
 W_{\pm \rm, rel} := \hbox{\rm s-lim}_{\pm \infty}\, e^{i\frac{t}{\hbar}H_{rel}}\,e^{ - i\frac{t}{\hbar}H_0, rel}.
 \ene
 The relative scattering operator,
 \beq\label{2.12}
 S_{\rm rel}= W_{+ \rm,  rel}^\ast \, W_{ -\rm, rel},
 \ene 
 is a unitary operator on ${\mathcal H}_{\rm rel}$.
 
We denote by $\hat{\mathcal H}:= L^2(\ere^6)$ the state space in the momentum representation.  The momentum of the particles  one and two are, respectively, $\p_1,\p_2$. We define the Fourier transform as an unitary operator from $\mathcal H$ onto $\hat{\mathcal H}$,

\beq\label{2.13}
\mathcal F \varphi(\p_1,\p_2):=\frac{1}{(2\pi \hbar)^{3}} \int_{\ere^6}\, e^{-\frac{i}{\hbar} (\p_1\cdot\x_1+\p_2\cdot \x_2) }\, \varphi(\x_1,\x_2).
\ene
It is also convenient to take as coordinates in the momentum representation the momentum of the center of mass and the relative momentum,
\begin{eqnarray}\label{2.14}  
 \p_{\rm cm}:= \p_1+\p_2, \\\label{2.14b}
\p:= \frac{m_2 \p_1- m_1\p_2}{m_1+m_2}.
\end{eqnarray}
The state space in the momentum representation also factorizes as a tensor product,
\beq\label{2.15}
\hat{\mathcal H}= \hat{\mathcal H}_{\rm cm} \otimes \hat{\mathcal H}_{\rm rel},
\ene
where $\hat{\mathcal H}_{\rm cm}=L^2(\ere^3), \hat{\mathcal H}_{\rm rel}:=L^2(\ere^3)$ are, respectively, the state spaces in the momentum representation for the center-of-mass motion and the relative motion.

The scattering operator in the momentum representation,
\beq\label{2.16}
\hat{S}:={ \mathcal F}\, S\, {\mathcal F}^{-1},
\ene  
decomposes as,
\beq\label{2.17}
\hat{S}= I_{\rm cm} \otimes \hat{S}_{\rm rel},
\ene
where $\hat{S}_{\rm rel}$ is the scattering operator for the relative motion in the momentum representation,
\beq\label{2.18}
\hat{S}_{\rm rel}:={ \mathcal F}_{\rm rel}\, S_{\rm rel}\, {\mathcal F}_{\rm rel}^{-1},
\ene
where ${ \mathcal F}_{\rm rel}$ is the Fourier transform in the relative coordinate,
   
\beq\label{2.19}
{\mathcal F}_{\rm rel} \varphi(\p):=\frac{1}{(2\pi \hbar)^{3/2}}\, \int_{\ere^3}\, e^{-\frac{i}{\hbar} \p\cdot\x }\, \varphi(\x).
\ene
We denote by $\ese^2$ the unit sphere in  $\ere^3$.

As $S_{\rm rel }$ commutes with $H_{\rm 0, rel}$ (energy conservation) we have 
\beq\label{2.20}
\left(\hat{S}_{\rm rel} \varphi \right)(\p)= \left({\mathcal S}(\p^2/2m)\varphi\right)(\p),
 \ene
 where the scattering matrix, ${\mathcal S}(E)$, is a unitary operator in $L^2(\ese^2)$ for each  $ E\in (0,\infty)$.
 Note that the scattering matrix defined in the time-dependent framework coincides with the scattering matrix defined in the stationary theory by means of the solutions to the Lippmann-Schwinger equations.
 
 The following theorem has been proved by Kato and Jensen \cite{kj}. Note that they consider the case $\hbar =1, m=1/2$ but the general case is easily obtained by an elementary argument. A zero energy resonance (half-bound state) is a solution to $H_{\mathrm rel}\varphi=0$ that decays at infinity, but that is not in $L^2(\ere^3)$. See \cite{kj} for a precise definition.  For generic potentials $V$ there  is neither a resonance nor an eigenvalue at zero for $H_{\mathrm rel}$. That is to say, if we consider the potential $\lambda V$ with a coupling constant $\lambda$, zero can be a resonance and/or an eigenvalue for at most a finite or denumerable set of $\lambda$'s without any finite accumulation point. 
 
 The scattering length is defined as,
 \beq \label{2.20b}
 c_0:= \frac{1}{ 4\pi}\,  \left(\frac{2m}{ \hbar^2} V\,(1+  G_0\frac{2m}{\hbar^2}  V)^{-1}1,1 \right),
\ene
where $(.,.)$ is the $L^2$ scalar product in $\ere^3$, $1$ designates the function identically equal to one and $G_0$ is the operator with integral kernel the Green's function at zero energy,
\beq\label{2.21}
G_0 (\x,\mathbf y):=  \frac{1}{ 4 \pi  |\x-\mathbf y| }, \x,\mathbf y \in \ere^3.  
\ene
The operator $(1+  G_0 \frac{2m}{\hbar^2} V)$ is invertible because zero is neither an eigenvalue nor a resonance for
 $H_{\mathrm rel}$. 
We define the scattering length, $c_0$, with the opposite sign to the one used in \cite{kj}, in order that it coincides with the definition used in the physics literature \cite{tay,ne}. Furthermore,

\beq \label{2.22}
Y_0(\nu):= \frac{1}{\sqrt{4\pi}}, \nu \in \ese^2,
\ene
\beq\label{2.23}
Y_1(\nu):= \frac{1}{4\pi^{3/2}}\, \left( \frac{2m}{ \hbar^2} V\,\left(1+ G_0  \frac{2m}{\hbar^2} V\right)^{-1}1, \x\cdot\nu   \right), \nu \in \ese^2.
\ene
We denote by ${\mathcal B}\left(L^2(\ese^2)\right)$ the Banach space of all bounded linear operators on $L^2(\ese^2)$.

\begin{theorem}(Kato and Jensen \cite{kj}) \label{th2.2}
Suposse that  Assumption \ref{ass2.1} is satisfied and that at zero $H_{\mathrm rel}$ has neither a resonance (half-bound state)  nor an eigenvalue.  Then, If $ \beta >5$, in the norm of ${\mathcal B}\left(L^2(\ese^2)\right)$ we have for $|\p/\hbar| \rightarrow 0$ the expansion,
\beq \label{2.24}
{\mathcal S}(\p^2/2m) = I + i |\p/\hbar| \, \Sigma^0_1 - |\p/\hbar|^2 \, \Sigma^0_2+ o(  |\p/\hbar|^2),
\ene
where $I$ is the identity operator on $L^2(\ese^2)$,
\beq \label{2.25}
\Sigma^0_1:= -2c_0 \left(\cdot,Y_0 \right) \,Y_0,
\ene
and
\beq \label{2.26}
\Sigma^0_2:= 2c_0^2 \left(\cdot,Y_0 \right) \,Y_0+ \left( \cdot, Y_1\right)  Y_0 - \left( \cdot, Y_0\right)  Y_1 .
\ene
Furthermore, if $\beta >7, o(|\p/\hbar|^2)$ can be replaced by $O(|\p/\hbar|^3)$.
\end{theorem}
Note that $Y_1=0$ if $V$ is spherically symmetric. We see that, as is well known, the leading order at low energy
of  ${\mathcal S}(\p^2/2m) - I$ is given by the scattering length, i.e. in leading order the scattering is isotropic. The anisotropic effects appear at second order. 

\section{Entanglement Creation}\sss
Consider a pure state of the two-particle system given in the momentum representation by the wave function $\varphi(\p_1,\p_2)$.  Let us denote by $\rho(\varphi)$ the one-particle reduced density matrix with integral kernel,
$$
\rho(\varphi)(\p_1,\p_1'):=\int \varphi(\p_1,\p_2)\,  \overline{ \varphi(\p_1',\p_2)}\,\, d \p_2,
$$

 and by ${\mathcal P}(\varphi)$ the purity,
\beq\label{3.1}
{\mathcal P}(\varphi):=\hbox{\rm Tr}(\rho^2)= \int_{\ere^{12}}\, d\p_1 \,d\p_1' \,d\p_2 \,d\p_2' \,\varphi(\p_1,\p_2) \,\overline{\varphi(\p_1', \p_2)} \,\varphi(\p_1',\p_2') \,\overline{\varphi(\p_1,\p_2')}.
\ene 
The purity is an entanglement measure that is closely related to the  R\'enyi entropy of
order $2$,$ - \hbox{\rm ln Tr}( \rho^2)$ \cite{ai, iko, lsa}. It is trivially related to the linear entropy, $S_L$, as $S_L=1-\mathcal P$.  It satisfies $ 0 \leq \mathcal P \leq 1$ if $\varphi$ is normalized to one. Furthermore, it is equal to one for a product state, $\varphi= \varphi_1(\p_1) \, \varphi_2(\p_2)$.  The purity is an entanglement measure that is convenient for the study of entanglement creation in scattering processes because it can be directly computed in terms of the scattering matrix.

We work in the  center-of-mass frame and we consider an incoming asymptotic state that is a product of two normalized Gaussian wave functions,	
\beq\label{3.2}
\varphi_{\rm in,\p_0}(\p_1,\p_2):=  \varphi_{\p_0}(\p_1) \,  \varphi_{-\p_0}(\p_2),
\ene
where
\beq\label{3.3}
\varphi_{\p_0}(\p_1):=  \frac{1}{(\sigma^2\pi)^{3/4}}   e^{- (\p_1-\p_0)^2 /  2\sigma^2}.
\ene
In the incoming asymptotic state (\ref{3.2}) particle one has mean momentum $ \p_0$ and particle two has mean momentum $ -\p_0$. The variance of the momentum distribution of both particles is $\sigma$.  We assume that the scattering takes place at the origin at time zero, and for this reason  the   average position of both particles  is zero in the incoming asymptotic state   (\ref{3.2}). Note that by (\ref{2.14b}) the mean value of the relative momentum in the state (\ref{3.2}) is equal to $\p_0$.

Since $\varphi_{\rm in, \p_0}$ is a product state its purity is one,
\beq\label{3.4b}
\mathcal P(\varphi_{\rm in,\p_0})=1.
\ene
After the scattering process is over the two particles are  in the outgoing  asymptotic state, $\varphi_{\rm out, \p_0}$, given by
 \beq\label{3.5}
 \varphi_{\rm out,\p_0}(\p_1,\p_2):= \left( \mathcal S( \p^2/2m ) \varphi_{\rm in, \p_0}\right)(\p_1,\p_2).
\ene
Since the relative momentum, $\p$, depends on $\p_1$ and on $\p_2$, $\varphi_{\rm out,\p_0}$ is no longer a product state and it has purity smaller than one, what means that entanglement between the two particles has been created by the scattering process. 

We will rigorously compute the leading order of the purity of $\varphi_{\rm out, \p_0}$ -in a quantitative way- in the 
low-energy limit for the relative motion. Note that to be in the low-energy regime  we need that the mean  relative momentum $\p_0$ be small, but also that the variance $\sigma$ be small, because if $\sigma$ is large the incoming asymptotic state   $\varphi_{\rm in, \p_0}$ will have a big probability of having large momentum, even if the mean relative momentum $\p_0$ is small.

We first introduce some notations that we need.

We  denote by $\varphi_{\rm in}$ the incoming asymptotic state  with mean value of the relative momentum zero,
\beq\label{3.6}
\varphi_{\rm in}(\p_1,\p_2):=  \varphi(\p_1) \,  \varphi(\p_2),
\ene
where,
\beq\label{3.7}
\varphi(\p):=  \frac{1}{(\sigma^2\pi)^{3/4}}   e^{-  \p^2 /  2\sigma^2}, \p \in \ere^3,
\ene
and by  $\varphi_{\rm out}$  the outgoing asymptotic state with incoming asymptotic state  $\varphi_{\rm in}$,
\beq\label{3.7b}
\varphi_{\rm out}(\p_1,\p_2):=  \left( \mathcal S( \p^2/2m ) \varphi_{\rm in}\right)(\p_1,\p_2).
 \ene
We define,
\beq\label{3.8}
\psi_{\q_0}(\q):=  \frac{1}{(\pi)^{3/4}}   e^{- (\q-\q_0)^2 /  2}, \q \in \ere^3,
\ene
\beq\label{3.9}
\psi(\q):=  \frac{1}{(\pi)^{3/4}}   e^{- \q^2 /  2},  \q \in \ere^3,
\ene
\beq\label{3.10}
\psi_{\rm in,\q_0}(\q_1,\q_2):=  \psi_{\q_0}(\q_1) \,  \psi_{-\q_0}(\q_2),
\ene
\beq\label{3.11}
\psi_{\rm in}(\q_1,\q_2):=  \psi(\q_1) \,  \psi(\q_2).
\ene

We prepare the following    proposition that we need  later.
\begin{prop} \label{prop3.1} 
\beq \label{3.12}
\left\|  \varphi_{\rm in,\p_0} -\varphi_{\rm in}\right\| \leq C \,   \hbox{\rm min} \{ |\p_0|/\sigma,1\},
\ene
\beq \label{3.13}
\left\| \p \left( \varphi_{\rm in,\p_0} -\varphi_{\rm in}\right)\right\| \leq C \, |\p_0|.
\ene
\end{prop}
\noindent {\it Proof:}  We denote $\q_0:= \p_0/\sigma.$ Then,
\beq \label{3.14}
\left\|\varphi_{\rm in,\p_0} -\varphi_{\rm in}\right\|= \left\|  \psi_{\rm in,\q_0} -\psi_{\rm in}\right\|.
\ene

Assume first that $|\q_0| \leq 1$. We have that,
\beq \label{3.16}
  \psi_{\rm in,\q_0} -\psi_{\rm in} = \frac{1}{\pi^{3/2}} \, e^{-(\q_1^2+\q_2^2)/2}  \left( e^{-\q_0^2+(\q_1-\q_2)\cdot \q_0} -1   \right).
  \ene
  Moreover,
  \begin{eqnarray}\label{3.17}
  \left| e^{-\q_0^2+(\q_1-\q_2)\cdot \q_0} -1 \right|= \left| \int_0^{-\q_0^2+(\q_1-\q_2)\cdot \q_0}\, e^s\, ds    \right| \leq
   \\\label{3.18}
   e^{ |\q_0|^2+ |\q_0|(|\q_1|+|\q_2|)}
 (|\q_0|^2+ |\q_0|(|\q_1|+|\q_2|)).
 \end{eqnarray}                 
   It follows from (\ref{3.14},\ref{3.16}, and \ref{3.18}) that (\ref{3.12}) holds for $\q_0| \leq 1$. In the case $|\q_0| \geq 1$ the estimate is immediate because,
   $$
   \left\|  \varphi_{\rm in,\p_0} -\varphi_{\rm in}\right\| \leq 2.
   $$
   Note that by (\ref{2.14b}),
   $$
   |\p|\leq |\p_1|+|\p_2|.
   $$
Then, if $|\q_0| \leq 1$ as in the proof of (\ref{3.12}) we prove  that,      
\beq \label{3.18b}
\left\| \p \left( \varphi_{\rm in,\p_0} -\varphi_{\rm in}\right)\right\| \leq C \, |\p_0|.
\ene
If $|\q_0| \geq 1$ we estimate as follows,
\beq\label{3.19}
\left\| \p \left( \varphi_{\rm in,\p_0} -\varphi_{\rm in}\right)\right\| \leq \left\| |\p_1| \left( \varphi_{\rm in,\p_0} -\varphi_{\rm in}\right)\right\|+\left\| |\p_2| \left( \varphi_{\rm in,\p_0} -\varphi_{\rm in}\right)\right\|.
\ene
Furthermore,

\begin{eqnarray}\label{3.20}
\left\| |\p_1|\, \left( \varphi_{\rm in,\p_0} -\varphi_{\rm in}\right)\right\| \leq  \left\| |\p_1-\p_0| \, \varphi_{\rm in,\p_0}\right\| +\left\| |\p_0| \, \varphi_{\rm in,\p_0} \right\|+\left\| |\p_1|\, \varphi_{\rm in}\right\|   \leq       \nonumber\\
\sigma \left\| |\q_1-\q_0| \, \psi_{\rm in,\q_0}\right\| + |\p_0| +\sigma    \left\| |\q_1|\, \psi_{\rm in}\right\|\ \leq  
  C |\p_0|.
\end{eqnarray}

In the last inequality we used that $\sigma \leq |\p_0|$.
In the same way we prove that,
\beq\label{3.21}
\left\| |\p_2| \left( \varphi_{\rm in,\p_0} -\varphi_{\rm in}\right)\right\|  \leq C |\p_0|.
\ene
By (\ref{3.19}, \ref{3.20}, \ref{3.21}) we have that,
\beq\label{3.22}
\left\| \p \left( \varphi_{\rm in,\p_0} -\varphi_{\rm in}\right)\right\| \leq    C  |\p_0|.
\ene
Equation (\ref{3.13})  follows from (\ref{3.18b}) and (\ref{3.22}).

\bull

Let us denote,
 \beq\label{3.23}
 \mathcal T(\p^2/2m):=\mathcal S(\p^2/m) -I,
 \ene
 where $I$ designates the identity operator on $L^2(\ese^2)$.
 It follows from (\ref{2.24}) and since $ \| \mathcal S(\p^2/2 m)\|_{\mathcal B(L^2(\ese^2))}=1$, that
 \beq\label{3.24}
 \left\| \mathcal T(\p^2/2 m) \right\|_{\mathcal B(L^2(\ese^2))}\leq C \frac{|\p/\hbar|}{1+|\p/\hbar|}.
 \ene
 Hence,
 \beq\label{3.25}
 \left\| \mathcal T(\p^2/2m)  \varphi_{\rm in}\right\| \leq C  \frac{\sigma}{\hbar}  \left\| \mathcal \q \, \psi_{\rm in}\right\|.
\ene
 
 We designate,
 \beq \label{3.26}
 \mathcal L(\phi_1,\phi_2,\phi_3,\phi_4):= \int_{\ere^{12}}\, d\p_1 \,d\p_1' \,d\p_2 \,d\p_2' \,\phi_1(\p_1,\p_2) \,\overline{\phi_2(\p_1', \p_2)} \,\phi_3(\p_1',\p_2') \,\overline{\phi_4(\p_1,\p_2')}.
\ene 
Note that,
$$
\mathcal P(\phi)= \mathcal L(\phi,\phi,\phi,\phi).
$$ 
 It follows from the Schwarz inequality that,
 \beq \label{3.27}
 \left| {\mathcal L} (\phi_1,\phi_2,\phi_3,\phi_4 ) \right | \leq \Pi_{j=1}^4 \|\phi_j \|.
 \ene
 The following theorem is  our first low-energy estimate of the purity.
 \begin{theorem}\label{theor3.2}
 Suposse that  Assumption \ref{ass2.1} is satisfied and that at zero $H_{\mathrm rel}$ has neither a resonance (half-bound state)  nor an eigenvalue. Then,  
 \beq\label{3.28}
 \mathcal P(\varphi_{\rm out, \p_0})= \mathcal P(\varphi_{\rm out})+O(|\p_0/\hbar|), \hbox{\rm as}\, |\p_0/\hbar| \rightarrow 0,
 \ene
 where  $O(|\p_0/\hbar|)$ is uniform on $\sigma$, for $\sigma$ in bounded sets.
\end{theorem} 
 \noindent{\it  Proof:}  Writing $\varphi_{\rm out,\p_0}$ as,
 $$ 
 \varphi_{\rm out,\p_0}:=  \mathcal S(\p^2/2m) \varphi_{\rm in,\p_0}=  \varphi_{\rm in,\p_0}+
  \mathcal T(\p^2/2m)\varphi_{\rm in,\p_0},
 $$
and using (\ref{3.4b}),  we see that we can write $\mathcal P(\varphi_{\rm out, \p_0})$ as follows,
 \beq\label{3.29}
 \mathcal P(\varphi_{\rm out, \p_0})= 1+ \mathcal R(\p_0),
 \ene
 where $\mathcal R(\p_0)$  is given by,
 \beq\label{3.30}
 \mathcal R(\p_0):= \sum_{i=1}^A\mathcal L_i(\p_0,\psi_1,\psi_2, \psi_3,\psi_4),
 \ene
for some  integer $A$, and  where each of the $\mathcal L_i(\p_0, \psi_1,\psi_2, \psi_3,\psi_4)$ is equal to,
\beq\label{3.31}
\mathcal L_i(\p_0, \psi_1,\psi_2, \psi_3,\psi_4)= \mathcal L(\psi_1,\psi_2, \psi_3,\psi_4),
\ene
where for some $ 1 \leq k \leq 4$,  $k$ of the   $\psi_j$  are equal to  $ \mathcal T(\p^2/2m) \varphi_{\rm in,\p_0}$ and the remaining $4-k$ are equal to $ \varphi_{\rm in, \p_0}$. Similarly,
\beq\label{3.32}
 \mathcal P(\varphi_{\rm out })= 1+ \mathcal R(0),
 \ene
 with
\beq\label{3.33}
 \mathcal R(0):= \sum_{i=1}^A\mathcal L_i(0,\psi_1,\psi_2, \psi_3,\psi_4).
 \ene
Below we prove that,
\beq\label{3.34}
  \mathcal R(\p_0)= \mathcal R(0)+O(|\p_0/\hbar|), \hbox{\rm as}\, |\p_0/\hbar| \rightarrow 0,
  \ene
  what proves the theorem in view of (\ref{3.29},\ref{3.32}).
  
  We proceed to prove (\ref{3.34}). Without losing generality we can assume that,
  \beq\label{3.35}
  \mathcal L_1(\p_0,\psi_1,\psi_2, \psi_3,\psi_4)= \mathcal L(\mathcal T(\p^2/2m) \varphi_{\rm in,\p_0},\varphi_{\rm in, \p_0},\varphi_{\rm in, \p_0},\varphi_{\rm in, \p_0}).
 \ene 
  We have that,
   \beq\label{3.36}
  \mathcal L_1(\p_0,\psi_1,\psi_2, \psi_3,\psi_4)= \mathcal L(\mathcal T(\p^2/2m) \varphi_{\rm in},\varphi_{\rm in, \p_0},\varphi_{\rm in, \p_0},\varphi_{\rm in, \p_0})+
  \mathcal L(\mathcal T(\p^2/2m) (\varphi_{\rm in,\p_0}-\varphi_{\rm in}),\varphi_{\rm in, \p_0},\varphi_{\rm in, \p_0},\varphi_{\rm in, \p_0}).
 \ene 
 By (\ref{3.13}, \ref{3.24}, \ref{3.27}, \ref{3.36}),
 \beq\label{3.37}
  \mathcal L_1(\p_0,\psi_1,\psi_2, \psi_3,\psi_4)=   \mathcal L(\mathcal T(\p^2/2m) \varphi_{\rm in},\varphi_{\rm in, \p_0},\varphi_{\rm in, \p_0},\varphi_{\rm in, \p_0})+ O(|\p_0/\hbar|), \hbox{\rm as}\, |\p_0/\hbar| \rightarrow 0.
  \ene
  In the same way, using (\ref{3.12}, \ref{3.25},\ref{3.37}), we prove that,
  \beq\label{3.38}
  \mathcal L_1(\p_0,\psi_1,\psi_2, \psi_3,\psi_4)=   \mathcal L(\mathcal T(\p^2/2m) \varphi_{\rm in},\varphi_{\rm in},\varphi_{\rm in, \p_0},\varphi_{\rm in, \p_0})+ O(|\p_0/\hbar|), \hbox{\rm as}\, |\p_0/\hbar| \rightarrow 0.
  \ene
  Repeating this argument two more times we obtain that,
    \begin{eqnarray}
  \mathcal L_1(\p_0,\psi_1,\psi_2, \psi_3,\psi_4)=   \mathcal L(\mathcal T(\p^2/2m) \varphi_{\rm in},\varphi_{\rm in},\varphi_{\rm in},\varphi_{\rm in, })+ O(|\p_0/\hbar|)=\nonumber
  \\
  \label{3.39}
  \mathcal L_1(0,\psi_1,\psi_2, \psi_3,\psi_4)+ O(|\p_0/\hbar|), \hbox{\rm as}\, |\p_0/\hbar| \rightarrow 0.
  \end{eqnarray}
We prove in the same way that,  
 \beq\label{3.40}
  \mathcal L_j(\p_0,\psi_1,\psi_2, \psi_3,\psi_4)=\mathcal L_j(0,\psi_1,\psi_2, \psi_3,\psi_4)+ O(|\p_0/\hbar|), 2\leq j\leq A, \hbox{\rm as}\, |\p_0/\hbar| \rightarrow 0.
  \ene
 Equation (\ref{3.34}) follows from (\ref{3.29},\ref{3.30},\ref{3.32},\ref{3.33}, \ref{3.39},\ref{3.40}).
  
  \bull

We now compute the leading order of the purity of  $\varphi_{\rm out }$.

We denote,
\beq \label{3.41}
\mathcal T_1( \p^2/2m):= {\mathcal S}(\p^2/2m) -I - i |\p/\hbar| \, \Sigma^0_1 + |\p/\hbar|^2 \, \Sigma^0_2.
\ene
It follows from Theorem  \ref{th2.2} that,
 \beq\label{3.42}
 \left\| \mathcal T_1(\p^2/m) \right\|_{\mathcal B(L^2(\ese^2))}\leq \left\{
 \begin{array}{clcr}
 |\p/\hbar|^2 o(1 ),    & \hbox{\rm if}\, \beta >5, \\
  |\p/\hbar|^2 O( |\p/\hbar|),   &   \hbox{\rm if}\, \beta >7,
  \end{array}
 \right.
 \ene
 where $o(1)$ and $ O( |\p/\hbar|)$ are bounded functions of $ |\p/\hbar|$, $\lim_{ |\p/\hbar| \rightarrow 0} o(1)=0$ and
 $O( |\p/\hbar|) \leq C  |\p/\hbar|$ for $ |\p/\hbar|\leq 1$.
 \begin{theorem}\label{theor3.3}
 Suposse that  Assumption \ref{ass2.1} is satisfied and that at zero $H_{\mathrm rel}$ has neither a resonance (half-bound state) nor an eigenvalue. Then, as $\sigma/\hbar $ goes to zero,  
 \beq\label{3.43}
 \mathcal P(\varphi_{\rm out})= \mathcal P\left(\left[I + i |\p/\hbar| \, \Sigma^0_1 - |\p/\hbar|^2 \, \Sigma^0_2\right]\varphi_{\rm in }\right)+\left\{
 \begin{array}{clcr}
o\left( |\sigma/\hbar|^2\right),    & \hbox{\rm if}\, \beta >5, \\
  O\left(|\sigma/\hbar|^3 \right),   &   \hbox{\rm if}\, \beta >7.
  \end{array}
 \right.
 \ene
 \end{theorem} 
 \noindent{\it  Proof:} We write  $\varphi_{\rm out}$ as follows,
 $$ 
 \varphi_{\rm out}=  \varphi_{\rm out, 1}+ \mathcal T_1(\p^2/2m)  \varphi_{\rm in},
$$
where,
\beq\label{3.44}
\varphi_{\rm out, 1}:= \left[I + i |\p/\hbar| \, \Sigma^0_1 - |\p/\hbar|^2 \, \Sigma^0_2\right]\varphi_{\rm in }. 
\ene
Using this decomposition we  write $\mathcal P(\varphi_{\rm out})$ as follows,
 \beq\label{3.45}
 \mathcal P(\varphi_{\rm out})= \mathcal P(\varphi_{\rm out, 1})+ \mathcal R(\sigma),
 \ene
 where $\mathcal R(\sigma)$  is given by,
 \beq\label{3.46}
 \mathcal R(\sigma):= \sum_{i=1}^B\mathcal L_i(\sigma,\psi_1,\psi_2, \psi_3,\psi_4),
 \ene
for some  integer $B$, and  where each of the $\mathcal L_i(\sigma, \psi_1,\psi_2, \psi_3,\psi_4)$ is equal to,
\beq\label{3.47}
\mathcal L_i(\sigma, \psi_1,\psi_2, \psi_3,\psi_4)= \mathcal L(\psi_1,\psi_2, \psi_3,\psi_4),
\ene
where for some $ 1 \leq k \leq 4$,  $k$ of the   $\psi_j$  are equal to  $\varphi_{\rm out,1}$ and the remaining $4-k$ are equal to $  \mathcal T_1(\p^2/2m)  \varphi_{\rm in}$. 

We proceed to prove that as  $\sigma/\hbar$ goes to zero,

\beq\label{3.48}
  \mathcal R(\sigma)= \left\{
 \begin{array}{clcr}
 o\left(|\sigma/\hbar|^2 \right),    & \hbox{\rm if}\, \beta >5, \\
  O\left(|\sigma/\hbar|^3 \right),   &   \hbox{\rm if}\, \beta >7,
  \end{array}
 \right.
\ene
  what proves the theorem in view of ( \ref{3.45}).
  
 Without any loss of generality we can assume that,
  \beq\label{3.49}
  \mathcal L_B(\sigma,\psi_1,\psi_2, \psi_3,\psi_4)= \mathcal L( \varphi_{\rm out, 1},  \varphi_{\rm out, 1}, \varphi_{\rm out, 1},\mathcal T_1(\p^2/2m)  \varphi_{\rm in}).
 \ene 
 By (\ref{3.27}, \ref{3.42})  We have that,
   \beq\label{3.50}
  \mathcal L_B(\sigma,\psi_1,\psi_2, \psi_3,\psi_4)= \left\{
 \begin{array}{clcr}
 o\left(|\sigma/\hbar|^2 \right),    & \hbox{\rm if}\, \beta >5, \\
  O\left(|\sigma/\hbar|^3\right),   &   \hbox{\rm if}\, \beta >7.
  \end{array}
 \right.
\ene 
We complete the proof of (\ref{3.48}) estimating the remaining terms in (\ref{3.46}) in the same way.

  \bull

We denote by 
\beq \label{3.51}
\mu_i= \frac{m_i}{m_1+m_2}, i=1,2,
\ene
the ratio of the mass of the $i$ particle to the total mass.

It follows from (\ref{2.14},\ref{2.14b}) that,
\beq \label{3.52}
\p_1= \mu_1 \p_{\rm cm}+\p,
\ene
\beq \label{3.53}
\p_2= \mu_2 \p_{\rm cm}-\p,
\ene
and that,
\beq\label{3.54}
\varphi_{\rm in}=  \frac{1}{(\sigma^2\pi)^{3/2 }} \,\,  \ds e^{ - \frac{\mu_1^2+\mu_2^2}{2\sigma^2} \p_{\rm cm}^2}    \, \,
\ds e^{ -\frac{\p^2+ (\mu_1-\mu_2) \p_{\rm cm}\cdot \p  }{ \sigma^2}}.
\ene
By  a simple computation using (\ref{3.44}, \ref{3.52}-\ref{3.54}) we prove that,
\beq\label{3.55}
\mathcal P\left( \left[I + i |\p/\hbar| \, \Sigma^0_1 - |\p/\hbar|^2 \, \Sigma^0_2\right]\varphi_{\rm in }\right)= 1- (\sigma/ \hbar)^2  \left( \mathcal  P_1(\psi_{\rm in})+\mathcal P_2(\psi_{\rm in})\right)+ O((\sigma/\hbar)^3), \quad \hbox{\rm as}\, \sigma/\hbar \rightarrow 0,
\ene
where,
\beq\label{3.56}
\mathcal P_1(\psi_{\rm in})= \Sigma_{j=1}^3 \mathcal P_{1,j}(\psi_{\rm in}),
\ene
with
\beq\label{3.57}
\mathcal P_{1,1}(\psi_{\rm in}) = -2 \int d\q_1d\q_2d\q_3\, |\mu_2\q_1-\mu_1\q_2| \,  |\mu_2\q_3-\mu_1\q_2|\,
(\Sigma^0_1\psi(\q_1,\q_2))\, (\Sigma^0_1\psi(\q_3,\q_2))\,\psi(\q_1,\q_3),
\ene
\beq\label{3.58}
\mathcal P_{1,2}(\psi_{\rm in})= -2 \int d\q_1d\q_2d\q_3\, |\mu_2\q_1-\mu_1\q_2| \,  |\mu_2\q_1-\mu_1\q_3|\,
\left(\Sigma^0_1\psi(\q_1,\q_2)\right)\, \left(\Sigma^0_1\psi(\q_1,\q_3)\right) \psi(\q_2,\q_3),
\ene
\beq\label{3.59}
\mathcal P_{1,3}(\psi_{\rm in})=2 \left[\int d\q_1d\q_2\, |\mu_2\q_1-\mu_1\q_2| \,
\left(\Sigma^0_1\psi(\q_1,\q_2)\right)\,  \psi(\q_1,\q_2)\right]^2,
\ene
and
\beq\label{3.60}
\mathcal P_{2}(\psi_{\rm in})=
4 \int d\q_1d\q_2\, |\mu_2\q_1-\mu_1\q_2|^2 \,
\left(\Sigma^0_2\psi(\q_1,\q_2)\right)\, \psi(\q_1,\q_2).
\ene
Explicitly evaluating the integrals in (\ref{3.57}), \ref{3.58},\ref{3.59})   using (\ref{3.54}) and $\mu_2=1-\mu_1$,  we prove that,
\beq\label{3.61}
\mathcal P_{1,1}(\psi_{\rm in}) = -8 \frac{c_0^2 \sigma^2}{\hbar^2}\, J(\mu_1,1-\mu_1),
\ene
\beq\label{3.62}
\mathcal P_{1,2}(\psi_{\rm in}) = -8 \frac{c_0^2 \sigma^2}{\hbar^2}\, J(1-\mu_1,\mu_1),
\ene
\beq\label{3.63}
\mathcal P_{1,3}(\psi_{\rm in}) = 8 \frac{c_0^2 \sigma^2}{\hbar^2}\,\left( L(\mu_1,1-\mu_1)\right)^2,
\ene
where,
\beq \label{3.64}\begin{array}{c}
J(\mu_1,\mu_2):= \frac{1}{\pi^{9/2}}\, \int d\q_2\left[ \,\,\int d\q_1 |\mu_2\q_1-\mu_1\q_2|\,\, \hbox{\rm Exp}[-\frac{1}{2}(\mu_1^2+\mu_2^2)(\q_1+\q_2)^2-(\mu_2\q_1-\mu_1\q_2)^2 - \q_1^2/2] \right.\\\\\left.
\ds \frac{\sinh[(\mu_1-\mu_2)\, |\q_1+\q_2|\,|\mu_2\q_1-\mu_1\q_2|]}{(\mu_1-\mu_2)\, |\q_1+\q_2|\,|\mu_2\q_1-\mu_1\q_2|}\,\,\right]^2,
 \end{array}
\ene
and
 \beq \label{3.65}\begin{array}{c}
L(\mu_1,\mu_2):=\frac{1}{\pi^3}\, \int d\q_1\, d\q_2 |\mu_2\q_1-\mu_1\q_2|\,\, \hbox{\rm Exp}[-(\mu_1^2+\mu_2^2)(\q_1+\q_2)^2-2(\mu_2\q_1-\mu_1\q_2)^2]
\\\\
 \hbox{\rm Exp}[ -(\mu_1-\mu_2)(\q_1+\q_2)\cdot (\mu_2\q_1-\mu_1\q_2)] 
\ds \frac{\sinh[(\mu_1-\mu_2)\, |\q_1+\q_2|\,|\mu_2\q_1-\mu_1\q_2|]}{(\mu_1-\mu_2)\, |\q_1+\q_2|\,|\mu_2\q_1-\mu_1\q_2|}.
 \end{array}
 \ene
 Furthermore,
 \beq\label{3.66}
 \mathcal P_{2}(\psi_{\rm in})= 8 \frac{c_0^2 \sigma^2}{\hbar^2}\, N(\mu_1,1-\mu_1),
\ene
where,
\beq\label{3.67}
 N(\mu_1,\mu_2):=  \frac{1}{\pi^3}\, \int d\q_{\rm cm}\, d\q \,\,\q^2 \, \hbox{\rm Exp}[-(\mu_1^2+\mu_2^2)\q_{\rm cm}^2-2\q^2-(\mu_1-\mu_2)\q_{\rm cm}\cdot \q]\, \frac{\sinh[(\mu_1-\mu_2)\, |\q_{\rm cm}|\,|\q|]}{(\mu_1-\mu_2)\,|\q_{\rm cm}|
 |\q|}.
\ene
Note that the  second and the third term in the right-hand side of (\ref{2.26}) give no contribution to  $ N(\mu_1,\mu_2)$ because as $Y_1(\nu)$ is an odd function the integrals of these terms are zero.  

By (\ref{3.43}, \ref{3.55}, \ref{3.56}, \ref{3.61}-\ref{3.63}, \ref{3.66})
\begin{eqnarray}\label{3.68}
 \mathcal P(\varphi_{\rm out})= 1-8 ( c_0\sigma/ \hbar)^2  \left((L(\mu_1,1-\mu_1))^2+N(\mu_1,1-\mu_1) -
J(\mu_1,1-\mu_1)-J(1-\mu_,\mu_1) \right)+  \nonumber \\   \left\{
 \begin{array}{clcr}
o\left( |\sigma/\hbar|^2\right),    & \hbox{\rm if}\, \beta >5, \\
 
 O\left(|\sigma/\hbar|^3 \right),   &   \hbox{\rm if}\, \beta >7.
  \end{array}
 \right.
 \end{eqnarray}

In the appendix we prove by explicit computation that,
\beq\label{3.69}
L(\mu_1,1-\mu_1)=\sqrt{ \frac{2}{\pi}}\, \left( 1+ (2\mu_1-1)^2 \right)^{-1/2},
 \ene
 \beq\label{3.70}
 N(\mu_1,1-\mu_1)= \frac{1}{2 (2\mu_1-1)^2}\, \frac{1}{\sqrt{1+(2\mu_1-1)^2}}\,\left[\left(1+(2\mu_1-1)^2\right)^{3/2}-1 \right],
 \ene
 \beq\label{3.71}
 N(1/2,1/2)= 3/4.
 \ene
We denote by $\mathcal E(\mu_1$) the entanglemement coefficient,
\beq\label{3.72}
\mathcal E(\mu_1):=8\left[( L(\mu_1,1-\mu_1))^2+N(\mu_1,1-\mu_1) -
J(\mu_1,1-\mu_1)-J(1-\mu_1,\mu_1)\right].
\ene
By (\ref{3.69},\ref{3.70}),
\beq\label{3.73}
\mathcal E(\mu_1):=  \frac{16}{\pi\, \left( 1+ (2\mu_1-1)^2 \right)}+ \frac{4}{ (2\mu_1-1)^2}\, \frac{\left(1+(2\mu_1-1)^2\right)^{3/2}-1 }{\sqrt{1+(2\mu_1-1)^2}}-8
J(\mu_1,1-\mu_1)-8 J(1-\mu_1,\mu_1).
\ene
Thus, we have proven the following theorem.
\begin{theorem}\label{theor3.6}
 Suposse that  Assumption \ref{ass2.1} is satisfied and that at zero $H_{\mathrm rel}$ has neither a resonance (half-bound state)  nor an eigenvalue.  Then, as $\sigma/\hbar$ goes to zero,
 \beq\label{3.74}
 \mathcal P(\varphi_{\rm out})=1- ( c_0\sigma/ \hbar)^2 \mathcal E(\mu_1) +\left\{
 \begin{array}{clcr}
o\left( |\sigma/\hbar|^2\right),    & \hbox{\rm if}\, \beta >5, \\
  O\left(|\sigma/\hbar|^3 \right),   &   \hbox{\rm if}\, \beta >7,
  \end{array}
 \right.
 \ene
 where the entanglement coefficient $\mathcal E(\mu_1)$ is given by (\ref{3.73}).
 \end{theorem} 
 
 \noindent {\it Proof:} The theorem follows from (\ref{3.68}, \ref{3.72}).
 
 \bull
 
 In the appendix we explicitly evaluate $J(1/2,1/2)$,
 \beq\label{3.75}
 J(1/2,1/2)= \frac{3}{2}+\frac{1}{\pi} \, \left(\frac{\sqrt{27}}{4}-3 \arctan\left(\frac{1}{2-\sqrt{3}}\right)\right) =0.663497.
\ene
By (\ref{3.73}, \ref{3.75}) for $\mu_1=1/2$, when the masses are equal, the entanglement coefficient is given by
\beq\label{3.76}
\mathcal E(1/2)= 0.4770.
\ene
We also explicitly evaluate in the appendix  $J(1,0)$,
\beq \label{3.77}
J(1,0)= 2(1+ \frac{1}{\sqrt{3}} -\sqrt{2})= 0.32627.
\ene 
For  $\mu_1 \in [0,1]\setminus\{1/2,1\}$ we compute  $J(\mu_1,1-\mu_1)$ numerically using Gaussian quadratures.

In Table 1 and in Figure 1 we give  values of $\mathcal E(\mu_1)$ for $0.5 \leq \mu_1:= m_1/(m_1+m_2) \leq 1$.

\begin{remark}\label{rem3.4}  {\rm Note that there is no term of order $\sigma/\hbar$ in (\ref{3.74}). Actually,  the terms
 of order  $\sigma/\hbar$ cancel each other because of the complex conjugates in the definition of the purity in (\ref{3.1})
  and of the factor $i$ in the second term in the right-hand side of (\ref{2.24})  that is there because of the unitarity of the scattering matrix. This shows that for low energy the entanglement is a second order effect.}
\end{remark}

\begin{remark}\label{rem3.5}
{\rm
Remark that $\mathcal E(\mu_1)=\mathcal E(1-\mu_1)$, what implies that the leading order in (\ref{3.74}) is invariant under the change $\mu_1 \leftrightarrow 1-\mu_1$, as it should be, because $\mathcal P(\varphi_{\rm out})$ is invariant under the exchange of particles one and two.}
\end{remark}
\begin{remark} \label{rem3.6}
{\rm 
As we mentioned in Remark \ref{rem3.4}, at low energy  the entanglement is a second order effect. As can be seen in 
Theorem \ref{th2.2},  for low energy the scattering is isotropic at first order and it is determined by the scattering length, $c_0$. However, the effects of the anisotropy of the potential appear at second order. It is quite remarkable that these effects give no contribution to the evaluation of $N(\mu_1,\mu_2)$,  as mentioned above. It follows from this that  the leading order of the entanglement for low energy  (\ref{3.74}) is determined by the scattering length, $c_0$,  and 
that  the anisotropy of the potential plays no role, on spite of the fact that entanglement is a second order effect, what is surprising. }
\end{remark}

\section{Conclusions} \sss

We considered  the entanglement creation in the low-energy scattering of two particles of mass $m_1,m_2$, in three dimensions with the interaction given by   potentials that are not required to be spherically symmetric. Initially the particles are in a pure state that is a product of two normalized Gaussian states with the same variance, $\sigma$, and opposite mean momentum.  The entanglement creation by the collision was measured by the purity, $\mathcal P$, of one of the particles in the state  after the collision. Before the collision the purity is one.

We gave  a rigorous computation, with error bound,   of the leading  order of the  purity, $ \mathcal P$, at  low-energy. Namely, we proved that the leading order of the purity is given by  $ 1- (c_0 \sigma/\hbar)^2 \mathcal E$, where $c_0$ is the scattering length and  the entanglement coefficient $\mathcal E$ depends only on the masses of the particles.

We proved that the entanglement  takes its minimum when the masses are equal and that it strongly increases with the differences of the masses. There is no term of order $\sigma/\hbar$ in  the leading order of the purity, what  shows that for low energy the entanglement is a second order effect. As is well known,  for low energy the effects of the anisotropy of the potential appear at second order.  It was found that these effects give no contribution to  the evaluation of the leading order of the purity and that  the anisotropy of the potential plays no role, on spite of the fact that entanglement  is a second order effect, what is surprising.

\section{Appendix} \sss
For the reader's convenience we compute on this appendix the integrals that we need in Section II.

We first state some elementary integrals that we need.
\beq\label{a.1}
\int_{0}^\infty e^{-ax^2}\, dx =\frac{1}{2}\,   \sqrt{\frac{\pi}{a}}, \quad a>0,
\ene

\beq\label{a.3}
\int_{0}^\infty e^{-ax^2}\, x^2\, dx=\frac{1}{4a} \sqrt{\frac{\pi}{a}},\,  \quad a>0,
\ene
\beq\label{a.4}
\int_{-\infty}^\infty e^{-ax^2-2bx}\, dx= \sqrt{\frac{\pi}{a}}\, e^{b^2/a}, \quad a>0,
\ene
\beq\label{a.5}
\int_{0}^\infty e^{-ax^2-2bx}\, dx= \frac{1}{2}\, \sqrt{\frac{\pi}{a}}\, e^{b^2/a}\, (1-\e(b/\sqrt{a})), \quad a>0,
\ene
where   $\e(x)$ is  the error function,
\beq  \label{a.6}
\e(x)= \frac{2}{\sqrt{\pi}}\, \int_0^x\, e^{-y^2}\, dy.
\ene

Equation  (\ref{a.3})  follows integrating by parts using $-(1/2a) \frac{\partial}{\partial x} e^{-ax^2}= e^{-ax^2}\, x$ and  (\ref{a.1}). Equations (\ref{a.4}, \ref{a.5}) follow from (\ref{a.1}) changing the variable of integration to $y= x+b/a$.
\beq\label{a.7}
\int  \frac{1}{(1+x^2)\,\sqrt{(2+x^2)}}\, dx= \arctan\left(\frac{1}{x}\right)+ \arctan\left(\frac{x}{2-\sqrt{2+x^2}}\right)+C.
\ene

We have that,
$$
\frac{\partial}{\partial z} (\e(\sqrt{z}))^2= \frac{4}{\pi}\, \int_0^1 dy\,   e^{-z(y^2+1)}\,.
$$
Hence,
$$
1- (\e(\sqrt{z}))^2= \int_z^\infty \,\frac{\partial}{\partial z} (\e(\sqrt{z}))^2 =  \frac{4}{\pi}\, \int_0^1\frac{ e^{-z(y^2+1)}}{y^2+1}\, dy.
 $$
 It follows that,
 \beq \label{a.8}
 (\e(z))^2= 1-  \frac{4}{\pi}\, \int_0^1\frac{ e^{-z^2(y^2+1)}}{y^2+1}\, dy.
\ene
 We first compute $J(1/2,1/2)$. By (\ref{3.64}),
 \beq\label{a.9}
 J(1/2,1/2)= \frac{1}{4 \pi^{9/2}}\, \int d\q_2  \, e^{-\q^2_2} g(|\q_2|)^2,
 \ene
 where,
 \beq \label{a.10}
 g(|\q_2|):= \int\, d\q_1|\q_1-\q_2|\, e^{-\q_1^2}\, d\q_1.
 \ene
To evaluate (\ref{a.10}) we take a system of coordinates where $\q_2=(|\q_2|,0,0)$, we  do the change of coordinates  
 $(q_{1,1}, q_{1,2},q_{1,3}) \rightarrow (q_{1,1}-|\q_2|, q_{1,2},q_{1,3})$ and we compute the integral in spherical coordinates, to obtain,
 \beq\label{a.11}
 g(|\q_2|)= \frac{ \pi\,e^{-|\q_2|^2}}{|\q_2|}\, \int_0^\infty \, e^{-\rho^2}\, \rho^2\,  (e^{2|\q_2| \rho}- e^{-2|\q_2| \rho}).
 \ene
 After repeated integrations by parts using   $-(1/2a) \frac{\partial}{\partial \rho} e^{-a\rho^2}= e^{-a\rho^2}\, \rho$ and  (\ref{a.5})
 we prove that,
 \beq\label{a.12}
  g(|\q_2|)= \frac{\pi^{3/2}}{2|\q_2|}\,  \e(|\q_2|) (1+ 2 |\q_2|^2)+ \pi e^{-|\q_2|^2}.
 \ene
 Introducing (\ref{a.12}) into (\ref{a.9}) and passing to spherical coordinates we obtain,
  \beq \label{a.13}
  J(1/2,1/2)= \frac{1}{\pi^{7/2}}\, \int_0^\infty \, \rho^2 \,e^{-\rho^2} \left(  \frac{\pi^{3/2}}{2\rho}\,  \e(\rho) (1+ 2 \rho^2)+
   \pi e^{-\rho^2}  \right)^2\, d\rho.
   \ene
 After expanding the square in the right-hand side of (\ref{a.13}), several integration by parts using 
     $-(1/2a) \frac{\partial }{\partial \rho} e^{-a\rho^2}= e^{-a\rho^2} \, \rho $ and (\ref{a.1}, \ref{a.3}, \ref{a.7}, \ref{a.8}) we obtain that,
 \beq \label{a.14}
  J(1/2,1/2)= \frac{3}{2}+\frac{1}{\pi} \, \left(\frac{\sqrt{27}}{4}-3 \arctan\left(\frac{1}{2-\sqrt{3}}\right)\right)=0.663497.
 \ene   
 We now compute $J(1,0)$. By (\ref{3.64})
 \beq\label{a.15}
 J(1,0)=  \frac{1}{4 \pi^{9/2}}\, \int d\q_2   \q_2^2\,   e^{-2\q_2^2}\, (h(\q_2))^2,
 \ene
 where,
 \beq
 \label{a.15b}
 h(\q_2)= \int d\q_1\,  e^{-(\q_1+\q_2)^2/2 }\, e^{-\q_1^2 /2}\,\, 
\frac{e^{|\q_1+\q_2|\,|\q_2|} -e^{-|\q_1+\q_2|\,|\q_2|} }{ |\q_1+\q_2|\,|\q_2|}.
\ene
 Changing the integration coordinate in (\ref{a.15b})  to $\bf Q=\q_1+\q_2$ we obtain,
 \beq \label{a.16}
 h(\q_2)=e^{-\q_2^2 /2} \, \int d\mathbf Q\,  e^{-(\mathbf Q)^2 }\, \, 
\frac{e^{|\mathbf Q|\,|\q_2|} -e^{-|\mathbf Q|\,|\q_2|} }{ |\mathbf Q|\,|\q_2|}\,    e^{\mathbf Q\cdot\q_2}.
\ene
 Using spherical coordinates and doing the integration in the angular variables we  get,
 \beq \label{a.17}
 h(\q_2)=\frac{2\pi}{|\q_2|^2}e^{-\q_2^2 /2} \, \int_0^\infty d\rho \,  e^{-\rho^2 }\,  
\left[e^{2 \rho\, |\q_2|} +e^{-2\rho\,|\q_2|} -2    \right] = \frac{2\pi}{|\q_2|^2}e^{-\q_2^2 /2} \, \int_{-\infty}^\infty d\rho \,  e^{-\rho^2 }\,  
\left[e^{2 \rho\, |\q_2|} -1  \right] .
\ene
By (\ref{a.1},\ref{a.4}),
\beq \label{a.18}
 h(\q_2)=\frac{2\pi^{3/2}}{|\q_2|^2}e^{-\q_2^2 /2} \,(e^{\q_2^2} -1 ).
 \ene
 Introducing (\ref{a.18}) into (\ref{a.15}) and performing the remaining integrals with the aid of (\ref{a.1} ) we prove that,
 \beq \label{a.19}
 J(1,0)= 2(1+ \frac{1}{\sqrt{3}} -\sqrt{2})= 0.32627.
  \ene   
 We proceed  to compute $L(\mu,1-\mu_1)$. Using spherical coordinates and performing the integrals in the angular variables we prove that for $\mu_1\neq \mu_2$,
 \beq\label{a.19b}
 L(\mu_1,\mu_2)= \frac{4}{(\mu_1-\mu_2)^2 \pi}\int_0^\infty \, d\lambda  \, \lambda\, e^{-2\lambda^2}\, \left[ \int_{-\infty}^\infty\,e^{-(\mu_1^2+\mu_2^2) \rho^2}  \left(e^{2(\mu_2-\mu_1) \lambda \rho} - 1 \right)\, d\rho \right] . 
 \ene
By  (\ref{a.1}, \ref{a.4}), and integrating by parts using  $-(1/2a) \frac{\partial}{\partial \rho} e^{-a\rho^2}= e^{-a\rho^2}\, \rho$ we prove that,
\beq\label{a.20}
L(\mu_1,1-\mu_1)=\sqrt{ \frac{2}{\pi}}\, \left( 1+ (2\mu_1-1)^2 \right)^{-1/2}.
 \ene
 Finally, we compute $N(\mu_1,1-\mu_1)$. Using spherical coordinates in (\ref{3.67}) and evaluating  the integrals in the angular coordinates we obtain for $\mu_1\neq \mu_2$ that,
 \beq\label{a.21}
 N(\mu_1,\mu_2)= \frac{4}{\pi (\mu_1-\mu_2)^2}\, \int_0^\infty\, d\lambda\, \lambda^2\, e^{-2\lambda^2}\, 
 \int_{-\infty}^\infty \, e^{-(\mu_1^2+\mu_2^2)\rho}\, \left( e^{2(\mu_1-\mu_2)\lambda\,\rho} -1 \right).
 \ene
 By (\ref{a.3},\ref{a.4}),
 \beq\label{a.22}
 N(\mu_1,1-\mu_1)= \frac{1}{2 (2\mu_1-1)^2}\, \frac{1}{\sqrt{1+(2\mu_1-1)^2}}\,\left[\left(1+(2\mu_1-1)^2\right)^{3/2}-1 \right].
 \ene
 Taking the limit as $\mu_1 \rightarrow 1/2$ we get,
 \beq
 N(1/2,1/2)= 3/4.
 \ene
 \noindent {\bf Acknowledgement}

\noindent I thank: Patrick Joly for his kind hospitality at the project POEMS, Institut Nationale de Recherche en Informatique et en Automatique Paris-Rocquencourt  where this  work was partially done, Gerardo Dar\'{\i}o Flores Luis and Sebastian Imperiale for their help in the numerical computation of $J(\mu_1,1-\mu_1)$ and Pablo Barbelis for usefull discussions on entanglement.

\newpage

\begin{table}\label{tab1}
 \caption{The Entanglement Coefficient $\mathcal E(\mu_1)$ } 
  \begin{center}
  \begin{tabular}{ l c  }
 $ \mu_1:= m_1/(m_1+m_2)$ & $\mathcal E (\mu_1)$  \\
  0.5 &  0.4770        \\
  0.525 &   0.4813              \\
  0.55    &   0.4937                      \\
  0.575    &      0.5144                  \\
   0.6   &       0.5434                    \\
   0.625   &     0.5816                \\
   0.65  &      0.6296               \\
    0.675  &        0.6880         \\
    0.7  &             0.7550                 \\
    0.725  &         0.8320                     \\
    0.75  &           0.9179                \\
    0.775  &        1.0120                  \\
    0.8  &             1.1130                \\
     0.825 &           1.2208           \\
     0.85 &               1.3228                 \\
     0.875 &           1.4488         \\
     0.9 &            1.5659              \\
     0.925 &          1.6832          \\
      0.95&        1.8010           \\
      0.975&            1.9168    \\
      1&              2.0287  
\end{tabular}
\end{center}
\end{table}
\begin{figure}\label{fig1}
\centering
\setlength{\unitlength}{1cm}
\includegraphics[width=25cm,totalheight=40cm]{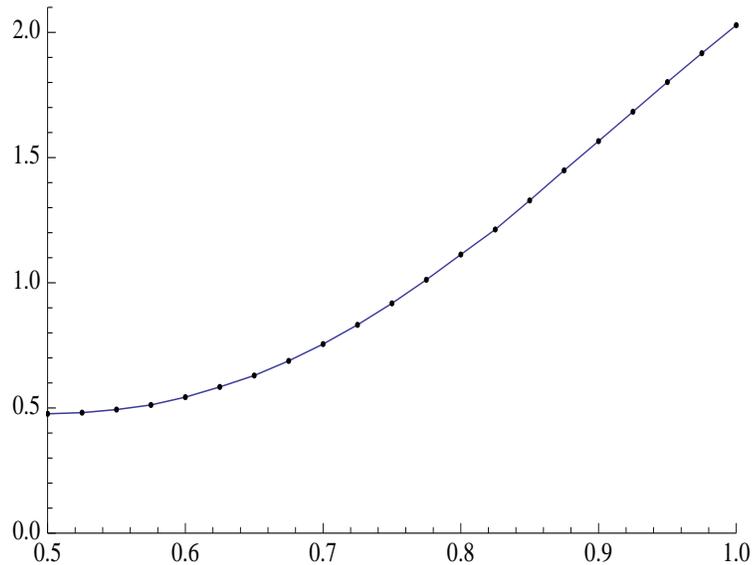}
\vspace{-29cm}
\caption{The entanglement coefficient  $y=\mathcal E(\mu_1)$, as a function of $x=\mu_1:= m_1/(m_1+m_2)$,  for $0.5\leq \mu_1 \leq 1.$}
\end{figure}
\end{document}